\documentclass[12pt]{iopart}
\usepackage{iopams}
\usepackage[utf8]{inputenc}
\usepackage[english]{babel}
\usepackage{cite}
\expandafter\let\csname equation*\endcsname\relax
\expandafter\let\csname endequation*\endcsname\relax 
\usepackage{amsfonts}
\usepackage{amssymb}
\usepackage{amsmath}
\usepackage{color}
\usepackage{mathrsfs}
\usepackage{relsize}
\usepackage{bm}
\usepackage{hyperref}
\usepackage{soul}
\usepackage{lineno}

\definecolor{dgreen}{rgb}{0.0, 0.2, 0.13}
\definecolor{orange}{rgb}{1.0, 0.27, 0.0}

\newcommand{\beq}{\begin{eqnarray}}
\newcommand{\eeq}{\end{eqnarray}}
\newcommand{\nn}{\nonumber}
\newcommand{\holst}{\bm{H}}

\def\keywords#1{\vspace{10pt}
     \begin{indented}
     \item[]\rm Keywords: #1\par
     \end{indented}}
     
\def\AMS#1{\vspace{10pt}
     \begin{indented}
     \item[]\rm AMS classification scheme numbers: #1\par
     \end{indented}}

\begin{document}


\title{Polysymplectic formulation for BF gravity with Immirzi parameter}
\author{Jasel Berra--Montiel$^{1,2}$, Alberto Molgado$^{1,2}$ and Ángel Rodríguez--López$^{1}$}

\address{$^{1}$ Facultad de Ciencias, Universidad Autonoma de San Luis 
Potosi \\
Campus Pedregal, 
Av.~Parque Chapultepec 1610,
Col.~Privadas del Pedregal,
San Luis Potosi, SLP, 78217, Mexico}
\address{$^{2}$ Dual CP Institute of High Energy Physics, Colima, Col, 28045, Mexico}

\eads{\mailto{\textcolor{blue}{jasel.berra@uaslp.mx}},\ 
\mailto{\textcolor{blue}{alberto.molgado@uaslp.mx}},\ \mailto{\textcolor{blue}{angelrodriguez@fc.uaslp.mx}}}

\begin{abstract}
The polysymplectic formulation of the CMPR action, which is a BF-type formulation of General Relativity that involves an arbitrary Immirzi parameter, is performed. We implement a particular scheme within this covariant Hamiltonian approach to analyze the constraints that characterize the CMPR model. By means of the privileged $(n-1)$-forms and the Poisson-Gerstenhaber bracket, inherent to the polysymplectic framework, the BF field equations associated to the CMPR action are obtained and,
in consequence, the Einstein equations 
naturally emerge by solving the simplicity constraints of the theory. Further, from the polysymplectic analysis of the CMPR action the De Donder-Weyl Hamiltonian formulation of the Holst action is recovered, which is consistent with the Lagrangian analysis of this model as reported in the literature.
\end{abstract}

\keywords{BF gravity, polysymplectic formalism, Poisson-Gerstenhaber bracket, Einstein equations, De Donder-Weyl theory, Immirzi parameter}

\pacs{04.20.Fy, 04.50.-h, 11.10.Ef, 11.15.Kc}

\AMS{83C05, 70S05, 70S15, 37K05}

\section{Introduction} 
Diffeomorphism invariant topological field theories are distinctively characterized by the 
absence of local degrees of freedom. 
These kind of gauge theories include the so-called 
BF theories which have been extensively analyzed in the literature showing a definitive strong relationship with Einstein theory of General Relativity~\cite{GHESDIT,Freidel2}. In particular, since the first BF-type formulation of General Relativity was introduced in~\cite{Pleb}, several extensions of the Plebanski action have been developed, 
giving rise to the denominated BF gravity models~\cite{Capo1,Krasnov,Freidel,Lewan,BFG,PAUGY}. Nowadays, the study of BF theories has been performed from different perspectives including  alternative formulations and models introduced 
with the aim to unify General Relativity and Yang-Mills theories as a formal starting point for the  spin foam approach for quantum gravity which is a non-perturbative quantization scheme for the gravitational field, as described for example 
in~\cite{BFG, FSMLQG, perez, baez}.
As it is well known, BF-type formulations of General Relativity are supplemented with 
extra conditions, known as the simplicity constraints, which set the explicit dependence of the $B$ field in terms of the tetrad, 
thus transforming the topological BF theory to the Palatini action for gravity~\cite{SCLQG}.
Of particular interest to us is a particular extension of the Plebanski action known as the CMPR action, introduced in~\cite{CMPR}. This action is described by a constrained BF gravity model that inserts a parameter which labels a family of canonical transformations on the phase space of General Relativity, and which within the 
Loop Quantum Gravity approach  has been shown to alter the spectra of certain geometrical quantities~\cite{IGRCCCG, ILQG}.  The aforementioned parameter is 
called the Immirzi parameter. 
The CMPR action is also useful to realize  equivalent formulations to some relevant BF gravity models commonly implemented within the spin foam formalism, as 
described in detail in~\cite{MMEBFG}. At the 
classical level, the CMPR action has been 
analyzed from different perspectives, including the Lagrangian formalism in~\cite{CMPR} and the Dirac-Hamiltonian approach in~\cite{LCHBGG} (see also~\cite{montesinos1,montesinos2} for more details on the CMPR model and BF theories), which strongly relies on the Hamiltonian analysis applied to the Plebanski action developed in~\cite{HHAPT, HANCPT}.  From our perspective, given the 
physical relevance and constraint structure 
of the model described by the CMPR action, it is a suitable candidate for its analysis under the geometric covariant Hamiltonian approach for field theories known as the polysymplectic formalism.

Based on the so-called De Donder-Weyl canonical theory introduced in~\cite{DDT, WT}, the polysymplectic framework starts with the definition of the polymomenta, which is a covariant extension of the standard momenta definition in the Hamiltonian approach for field theories.  These polymomenta include information on the variation of the action with
respect to all of the spacetime derivatives of the configuration variables, and induce the polymomenta phase-space~\cite{IKCSCHF,kana2}.  
The polymomenta phase-space is endowed with a canonical $(n+1)$-form, known as the polysymplectic form, which encodes the relevant physical data of a classical field theory in the sense that this $(n+1)$-form
contains sufficient information in order to  construct a graded Poisson-Gerstenhaber bracket in the space of differential forms~\cite{IKCSCHF, IKCSPPS, IKGPA}. The introduction of this Poisson-Gerstenhaber bracket to the polysymplectic formalism allows us to identify in turn a particular set of canonical conjugate 
$(n-1)$-forms, which play the role of canonical variables within this framework, for which the De Donder-Weyl-Hamilton covariant equations may be deduced.  
It is important to mention that there are some other covariant geometric 
approaches closely related to the De Donder-Weyl theory, such as the 
multisymplectic formalism in which one is able to recover the right equations 
from the multisymplectic form and also one may construct an alternative Poisson bracket 
for $(n-1)$-forms.  However, contrary to the situation within the polysymplectic formalism,  the Poisson bracket constructed in the multisymplectic framework results too restrictive to reproduce the algebra of 
observables of a given field theory and it results not suitable to obtain the right field equations, as discussed in~\cite{IKCSPPS,Forger2}.  
The Poisson-Gerstenhaber bracket structure inherent to the polysymplectic formalism together with the canonical conjugate $(n-1)$-forms thus play an important role in order to implement either a pre-canonical quantization scheme as described in~\cite{IKHEQFT, IKQFTPV, IKPSWF,kana1,kana2} or a deformation quantization for a field theory~\cite{david}.
Some physically motivated examples for which 
the polysymplectic framework has been applied and developed in detail may be 
encountered in references~\cite{IKCSCHF, MMCF, IKHFVG, DVVG, IKPQYMT, JEA, JAD,kana3}.

Our main purpose in this paper is thus to obtain a consistent polysymplectic formulation of the 
CMPR action. To this end, we will implement the proposal to analyze singular Lagrangian systems within the polysymplectic approach as introduced in~\cite{IKGD}, and which consists of a covariant extension of Dirac's formalism for constrained systems. 
Contrary to the standard Hamiltonian formulation of the CMPR action, we find that within the polysymplectic framework the analogous full set of constraints that characterizes the CMPR action is second-class in Dirac's terminology and, in consequence, the correct field equations of the system are obtained by means of the De Donder-Weyl-Hamilton equations on the constraint surface, that is, the surface in polymomenta phase space where the constraints are strongly satisfied as identities. In particular, the so-called simplicity 
constraints of the CMPR model are obtained straightforwardly and, after substituting the solution of these constraints into the polysymplectic formulation of the CMPR action, 
either the Palatini or the Holst actions for General Relativity are recovered.  This issue is clearly consistent with the Lagrangian analysis of the CMPR action as reported in the literature. 
However, we must emphasize that the manifestly covariant polysymplectic formalism allowed us to recover  Einstein equations in terms of the tetrads
directly from the BF action and without recursively introducing the Holst action.  Besides, as the polysymplectic phase space is finite dimensional,
contrary to the standard canonical Hamiltonian formalism where the field  variables are defined on some space-like hypersurface thus implying an infinite dimensional phase space,  
our formalism may pave the way to analyze novel quantum aspects of 
General Relativity thought as a BF theory from the perspective of 
an explicitly covariant canonical formulation~\cite{kana2,IKHFVG,kana3}.


The rest of the paper is organized as follows: in Section~\ref{PF} we give a brief introduction to the polysymplectic formalism for classical field theory in order to consider all the required background and introduce our set of notations.  We also include a concise discussion of the analysis of singular Lagrangian systems within this approach. In Section~\ref{PAM} we briefly describe the CMPR model for gravity following the notation introduced in~\cite{MMEBFG}, and we present the polysymplectic analysis of the CMPR model, 
obtaining its associated field equations which are equivalent to Einstein equations, after solving the simplicity constraints of the theory. Besides, we also discuss the way in which the CMPR action reduces into the Holst action at the polysymplectic level. Finally, in Section~\ref{SC} we introduce some concluding remarks.  

\section{Polysymplectic formalism}\label{PF}
In this section, we briefly introduce the polysymplectic formulation for classical field theories. We will closely follow the description of the geometric and algebraic structures of the formalism as described in \cite{IKCSCHF, IKCSPPS, IKGPA, IKHEQFT, IKQFTPV, IKPSWF}. We will also include a summary of the proposal to analyze singular Lagrangian systems within the polysymplectic approach as described in~\cite{IKGD}.

To start, we will consider a fibre bundle $(E, \pi, \mathcal{M})$, where $E$ is the total space whose fibers are $m-dimensional$ smooth
manifolds with local coordinates $\{y^{a}\}_{a=0}^{m-1}$, and $\mathcal{M}$ is the denominated base space which is an $n$-dimensional smooth manifold with local coordinates $\{x^{\mu}\}_{\mu=0}^{n-1}$, and will be identified with the spacetime manifold.  Finally,  $\pi: E\rightarrow \mathcal{M}$ is the standard projector map associated to the fiber bundle. Let $\phi:\mathcal{M}\rightarrow E$ be a section defined at a point $p\in\mathcal{M}$ whose local representation on $E$ is given by the composition $\phi^{a}:=y^{a}\circ\phi$. The set of all sections of $\pi$ at a point $p\in\mathcal{M}$ will be denoted by $\Gamma_{p}(\pi)$. Thus, we define the first jet manifold $J^{1}E$ of $(E, \pi, \mathcal{M})$, whose elements $j_{p}^{1}\phi\in J^{1}E$ have the local coordinate representation $(x^{\mu}, \phi^{a}, \phi^{a}_{\mu})$ and stands for the configuration space of the theory, where $\phi^{a}_{\mu}:=\partial \phi^{a}/\partial x^{\mu}$ denotes the field derivatives with respect to spacetime coordinates \cite{GJB, GLFT}.

Now, let us consider $\mathcal{U}\subset \mathcal{M}$ an open  submanifold of the base space $\mathcal{M}$ where, given an explicit  Lagrangian density $\mathcal{L}:J^{1}E\rightarrow\mathbb{R}$, a field theory can be defined by the action
\beq
\label{FL}
\mathcal{S}[\phi]:=\int_{\mathcal{U}}\mathcal{L}(j^{1}_{p}\phi)\, \omega\, ,
\eeq
where $\omega:=dx^{0}\wedge \cdots \wedge dx^{n-1}$ denotes the local volume $n$-form element of $\mathcal{M}$. In order to develop a covariant Hamiltonian formulation of the field theory  \eqref{FL}, we introduce a new set of variables, called polymomenta, and given by \cite{IKCSCHF}
\beq
\label{PMD}
\pi^{\mu}_{a}:=\frac{\partial \mathcal{L}}{\partial \phi^{a}_{\mu}}\, ,
\eeq
which induce an associated affine dual bundle of the jet manifold $J^{1}E$ that we will denominate as the polymomenta phase-space, 
and may be regarded as a smooth manifold, denoted by $\mathcal{P}$, whose local coordinate representation is given by $z^{M}:=(x^{\mu}, \phi^{a}, \pi^{a}_{\mu})$, as described in~\cite{IKCSPPS}. Thus, by means of the De Donder-Weyl-Legendre transformation $\mathbb{F}L:\mathbf{C}^{\infty}(J^{1}E)\rightarrow \mathbf{C}^{\infty}(\mathcal{P})$, the Lagrangian density, $\mathcal{L}\in\mathbf{C}^{\infty}(J^{1}E)$, is associated to the denominated \textit{De Donder-Weyl Hamiltonian}, $H_{\mathrm{DW}}\in \mathbf{C}^{\infty}(\mathcal{P})$, the latter
being locally represented as
\beq
\label{DWHD}
H_{\mathrm{DW}}(x^{\mu}, \phi^{a}, \pi^{a}_{\mu}):=\mathbb{F}L(\mathcal{L})=\pi^{\mu}_{a}\phi^{a}_{\mu}-\mathcal{L}(x^{\mu},\phi^{a},\phi^{a}_{\mu}),
\eeq
which together with the polymomenta \eqref{PMD} are central to the so-called De Donder-Weyl canonical theory which appeared for the first time in \cite{DDT, WT}.

In order to follow the geometrical description of the polysymplectic formalism, we will split the polymomenta phase-space, $z^{M}=(x^{\mu},z^{v})$, into the horizontal and vertical subspaces which are locally  represented by $(x^{\mu})$ and $z^{v}:=(\phi^{a},\pi_{a}^{\mu})$, respectively (see \cite{IKCSCHF, IKCSPPS} for further details and discussion). Given the space of all sections of the tangent bundle $T\mathcal{P}$, denoted by $\mathfrak{X}(\mathcal{P})$, we will term the set of vectors that does not have tangent components on the base space manifold $\mathcal{M}$ by $\mathfrak{X}^{V}(\mathcal{P})$.  Those vectors are sections of $T^{V}\mathcal{P}$, that is, the vertical tangent bundle of $\mathcal{P}$. Thus, given $X\in\mathfrak{X}^{V}(\mathcal{P})$ its local representation is defined by $X=X^{v}\partial_{v}:=X^{a}\partial_{a}+X^{\mu}_{a}\partial^{a}_{\mu}$, where we have adopted the short notation $\partial_{a}:=\partial/\partial \phi^{a}$ and $\partial^{\mu}_{a}:=\partial/\partial \pi^{\mu}_{a}$. 

Now, given the space of $r$-forms $\Omega^{r}(\mathcal{P}):=\bigwedge^{r}\Omega^{1}(\mathcal{P})$, that is, sections of $\bigwedge^{r}T ^{*}\mathcal{P}$, we will denote the space of horizontal $(r;s)$-forms by $\Omega^{r}_{s}(\mathcal{P})\in \Omega^{r}(\mathcal{P})$ as the set of elements $\theta$ satisfying $X\lrcorner\,\theta\in \Omega^{r-1}_{s-1}(\mathcal{P})$ for $X\in \mathfrak{X}^{V}(\mathcal{P})$, where $s$ denotes the degree of the differential forms on the vertical subspace of $\mathcal{P}$.

The polymomentum phase-space is endowed with a canonical $n$-form, $\Theta_{\mathrm{PC}}$, known as the Poincaré-Cartan form, given by \cite{GJB, MMCF}
\beq
\label{PCF}
\Theta_{\mathrm{PC}}:=\pi^{\mu}_{a}d\phi^{a}\wedge\omega_{\mu}-H_{\mathrm{DW}}\, \omega\, ,
\eeq
where $\omega_{\mu}:=\partial_{\mu}\lrcorner\,\omega$. Thus, the exterior derivative of $\Theta_{\mathrm{PC}}$ induces, in the polymomenta phase-space, the canonical $(n+1)$-form   
\beq
\label{MF}
\Omega_{\mathrm{DW}}=d\pi^{\mu}_{a}\wedge d\phi^{a}\wedge\omega_{\mu}-dH_{\mathrm{DW}}\wedge \omega\, ,
\eeq
which is known as the De Donder-Weyl multisymplectic form, and it contains all the information about the  classical dynamics of the fields, as we will see below. However, as described 
in~\cite{IKCSCHF}, it is possible to construct a well-defined 
Poisson structure if one only                considers the vertical term of the Poincaré-Cartan form, that is,  $\Theta^{V}_{\mathrm{PC}}:=\pi^{\mu}_{a}d\phi^{a}\wedge\omega_{\mu}$. 
By means of  this Poisson structure one may induce the De Donder-Weyl-Hamilton equations for a physical system.
In consequence, we will only consider the vertical Poincaré-Cartan form $\Theta^{V}_{\mathrm{PC}}$ to construct the geometric and algebraic structures in order to describe the dynamics of physical systems within the De Donder-Weyl field theory.
This is the basis for the polysymplectic formalism~\cite{IKCSCHF}. Thus, we may define the vertical exterior derivative, $d^{V}$,  acting on an arbitrary $p$-form $\Phi$ as
\beq
\label{VDD}
d^{V}\Phi:=\frac{1}{p!}\partial_{v}\Phi_{M_{1}\cdots M_{p}}dz^{v}\wedge dz^{M_{1}}\wedge \cdots\wedge dz^{M_{p}}\, ,
\eeq
the vertical part of the Poincaré-Cartan form $\Theta^{V}_{\mathrm{PC}}$ induces a canonical $(n+1;2)$-form on the polymomenta phase-space
\beq
\Omega^{V}_{\mathrm{DW}}:=d\pi^{\mu}_{a}\wedge d\phi^{a}\wedge\omega_{\mu}\, ,
\eeq
which is simple known as the polysymplectic 
De Donder-Weyl form \cite{IKCSCHF, IKCSPPS, IKGPA}. 
The vertical exterior derivative has been explored extensively in the literature, and we encourage the reader to check~\cite{CPVD} for further mathematical properties.
We define a vertical $p$-multivector field as a vector field that has one vertical and $(p-1)$ horizontal indices, namely $\overset{p}{X}=\frac{1}{p!}X ^{v\mu_{1}\cdots\mu_{p-1}}\partial_{v}\wedge\partial_{\mu_{1}}\wedge\cdots\wedge\partial_{\mu_{p-1}}$. 
We will use this type of $p$-multivector fields
in order to construct the analogous to the 
Hamiltonian vector fields within this context.
In particular, for a vertical $p$-multivector field $\overset{p}{X}_{F}$  we can assign a unique horizontal $(n-p;0)$-form $\overset{n-p}{F}$ by means of 
\beq
\label{HMFD}
\overset{p}{X}_{F}\lrcorner\, \Omega^{V}_{\mathrm{DW}}=d^{V}\overset{n-p}{F}\, .
\eeq
Thus, we say that $\overset{p}{X}_{F}$ is a Hamiltonian $p$-multivector field and $\overset{n-p}{F}$ its associated Hamiltonian horizontal $(n-p;0)$-form \cite{IKCSPPS}.

In addition, for a pair of the Hamiltonian forms $\overset{p}{F}\in\Omega^{p}_{0}(\mathcal{P})$ and $\overset{q}{G}\in\Omega^{q}_{0}(\mathcal{P})$, given their associated Hamiltonian multivector fields $\overset{n-p}{X_{F}}$ and $\overset{n-q}{X_{G}}$. The map $\{\![\,\cdot, \cdot\,]\!\}: \Omega^{p}_{0}(\mathcal{P})\times \Omega^{q}_{0}(\mathcal{P})\rightarrow \Omega^{p+q-n+1}_{0}(\mathcal{P})$ given by
\beq
\label{PGD}
\{\![\overset{p}{F}, \overset{q}{G}]\!\}:=(-1)^{n-p}\overset{n-p}{X_{F}}\lrcorner \overset{n-q}{X_{G}}\lrcorner \, \Omega^{V}_{\mathrm{DW}}\, ,
\eeq
is called the Poisson-Gerstenhaber bracket, which by definition is only defined for $p+q \geqslant n-1$ and obeys the following graded-commutation relation $\{\![\overset{p}{F}, \overset{q}{G}]\!\}=-(-1)^{|F||G|}\{\![\overset{q}{G}, \overset{p}{F}]\!\}$, where $|F|:=n-p-1$ and $|G|:=n-q-1$ are the degrees of the horizontal forms $\overset{p}{F}$ and $\overset{q}{G}$ with respect to the Poisson-Gerstenhaber bracket, respectively. One may also check that this bracket fulfills the graded Jacobi identity $\{\![\overset{p}{F}, \{\![\overset{q}{G}, \overset{r}{H}]\!\}]\!\}=\{\![\{\![\overset{p}{F}, \overset{q}{G}]\!\}, \overset{r}{H}]\!\}+(-1)^{|F||G|}\{\![\overset{q}{G}, \{\![\overset{p}{F}, \overset{r}{H}]\!\}]\!\}$. Finally, this bracket satisfies a graded Leibniz rule
\beq
\{\![\overset{p}{F},\overset{q}{G}\bullet\overset{r}{H}]\!\}=\{\![\overset{p}{F}, \overset{q}{G}]\!\}\bullet\overset{r}{H}+(-1)^{(n-q)|F|}\overset{q}{G}\bullet\{\![\overset{p}{F}, \overset{r}{H}]\!\}\, ,
\eeq
where the map $\bullet: \Omega^{p}_{0}(\mathcal{P})\times \Omega^{q}_{0}(\mathcal{P})\rightarrow \Omega^{p+q-n}_{0}(\mathcal{P})$, denominated as the co-exterior product, is given by
\beq
\label{CPD}
\overset{p}{F}\bullet\overset{q}{G}:=*^{-1}\left( *\overset{p}{F}\wedge * \overset{q}{G} \right)\, .
\eeq
Here the $*$~symbol stands for the Hodge dual operator only defined on the base space $\mathcal{M}$ \cite{IKGPA, IKHEQFT}. In addition, we also note from the definition of the Poisson-Gerstenhaber bracket~\eqref{PGD} that the Hamiltonian $(n-1)$-forms will play a primordial role within the polysymplectic formalism as the $(n-1)$-forms close under this bracket
structure. From the physical point of view, the relevance of the $(n-1)$-forms is associated to the fact that it is possible construct Noether currents 
with them, and these currents will serve to 
obtain certain physical observables for a given field theory~\cite{Helein,Forger2,khavkine}. In particular, it is possible to induce canonically conjugate variables considered as $(n-1)$-forms by the Poisson-Gerstenhaber bracket 
\beq
\label{CRD}
\{\![\pi^{\mu}_{a}\omega_{\mu}, \phi^{b}\omega_{\nu}]\!\}=\delta^{b}_{a}\omega_{\nu}\, ,
\eeq 
where, as before, $\omega_{\mu}:=\partial_{\mu}\lrcorner\,\omega$ denote the basis for the 
horizontal $(n-1)$-forms.  These relations also play an important role in order to implement a precanonical quantization of this formalism as described in \cite{IKHEQFT, IKQFTPV, IKPSWF}.
As discussed in~\cite{IKCSPPS}, there exists a relation of the 
Gerstenhaber bracket 
constructed within the polysymplectic framework with the conventional 
equal-time bracket within the Hamiltonian formalism for fields 
given by 
\begin{equation}
\int_{\Sigma}\int_{\Sigma}
\{\pi^0_a(\mathbf{x}_1),\phi^b(\mathbf{x}_2)\}_{PB} f(\mathbf{x}_1)g(\mathbf{x}_2)d\mathbf{x}_1d\mathbf{x}_2=
\int_\Sigma \{\![\pi_a, \phi^{b}\omega_{0}]\!\}f(\mathbf{x}_1)g(\mathbf{x}_1)  \,,
\end{equation}
where $\Sigma$ stands for a space-like Cauchy surface in the base spacetime manifold $\mathcal{M}$, $\mathbf{x}_1,\mathbf{x}_2\in\Sigma$, $f(\mathbf{x})$ and 
$g(\mathbf{x})$ are test functions on $\Sigma$ and, on the left-hand side we have
introduced the projection of the polymomentum $\pi_a$ into $\Sigma$, that is,
$\pi_a|_\Sigma=\pi^0_a(\mathbf{x},t)\omega_0$, being $t:=x^0$.

The co-exterior product~(\ref{CPD}) induces a derivative operator, $\mathbf{d}\bullet:
\Omega^p_0(\mathcal{P})\rightarrow \Omega^{p-(n-1)}_0(\mathcal{P})$, called the co-exterior derivative which, for an horizontal $(n-1)$-form 
$\overset{n-1}{F}=F^{\mu}\omega_{\mu}$, locally can be represented by
\beq
\label{DCD}
\mathbf{d}\bullet \overset{n-1}{F}:=\partial_{\nu}F^{\mu}\partial_{\rho}z^{\nu}\,dx^{\rho}\bullet \omega_{\mu}+\mathbf{d}^h\bullet \overset{n-1}{F}\, ,
\eeq
where the last term is called the horizontal co-exterior derivative and locally reads
$
\mathbf{d}^h\bullet\overset{n-1}{F}:=\partial_{\rho}F^{\mu}dx^{\rho}\bullet\omega_{\mu}
$.  Note that the co-exterior derivative is only defined for $(n-1)$ and $n$-forms.
Altogether, considering the De Donder-Weyl Hamiltonian \eqref{DWHD}, the Poisson-Gerstenhaber bracket \eqref{PGD} and the co-exterior product definition \eqref{CPD} we may write the co-exterior derivative of a Hamiltonian $(n-1)$-form $\overset{n-1}{F}$  in terms of the Poisson-Gerstenhaber bracket as
\beq
\label{eq:evolution}
\mathbf{d}\bullet\overset{n-1}{F}=-\sigma(-1)^{n}\{\![H_{\mathrm{DW}},\overset{n-1}{F}]\!\}+\mathbf{d}^h\bullet \overset{n-1}{F}\, ,
\eeq
where $\sigma=\pm 1$ depends on the signature  of the metric of the base space manifold $\mathcal{M}$ \cite{IKHEQFT}.  The last relation allows us to write the De Donder-Weyl-Hamilton equations as 
\beq
\nn
\partial_{\mu}\pi^{\mu}_{a}
& = &
\{\![H_{\mathrm{DW}},\pi^{\mu}_{a}\omega_{\mu}]\!\}=- \partial_{a}H_{\mathrm{DW}}\, , \nn\\
\partial_{\mu}\phi^{a}
& = & 
\{\![H_{\mathrm{DW}},\phi^{a}\omega_{\mu}]\!\}=\partial^{a}_{\mu}H_{\mathrm{DW}}\, .
\label{DWHED}
\eeq
One may straightforwardly show, by considering the definition of the 
polymomenta~\eqref{PMD} and the De Donder-Weyl Hamiltonian~\eqref{DWHD}, that these equations are completely equivalent to Euler-Lagrange field equations whenever the Lagrangian density is non-singular, that is, in the case the determinant of the Hessian matrix is non-vanishing, $\det\left(\partial^{2}\mathcal{L}/\partial \phi_{\mu}^{a}\partial \phi_{\nu}^{b}\right)\neq 0$. On the contrary, if 
this determinant does vanish the Lagrangian 
is called singular, and in this case 
we are not able to invert the gradients of the field variables $\phi^{a}_{\mu}$ in terms of the field variables $\phi^{a}$ and the polymomenta $\pi^{\mu}_{a}$. Thus, for singular systems we obtain a set of conditions defining a surface 
in the polymomenta phase-space, $\mathcal{P}$,
where the  Legendre transformation is invertible.  
Those conditions emerge from the definition of the 
polymomenta~\eqref{PMD}, namely,
\beq
\label{PCD}
\mathcal{C}_{k}^{(1)\mu}(\phi^{a},\pi^{\nu}_{a})\approx 0\, , \hspace{5ex}  
k=1,\ldots,j\leq mn  \,,
\eeq 
which are denominated primary constraints in Dirac's terminology~\cite{QSQ}. 
We would like to emphasize that, even though Dirac's terminology 
stands for an abuse of language, it will be very convenient in order to 
analyze the field equations for the model of our interest.
 Here the weak equality symbol, $\approx$, 
means that it is evaluated at the constrained surface, as in the standard Dirac's approach for constrained systems.  Also, the 
subindex $k$ labels each of the $j$ primary constraints.    From now on, instead of using these set of constraints, we will consider the 
associated constraint $(n-1)$-form $\mathcal{C}_{k}^{(1)}:={\mathcal{C}_{k}}^{(1)\mu}\omega_{\mu}$, and 
in analogy to Dirac formalism for singular Lagrangian systems, we define the total De Donder-Weyl Hamiltonian, $\widetilde{H}_{\mathrm{DW}}$, as the De Donder-Weyl Hamiltonian subject to the primary constraints \eqref{PCD} 
\beq
\label{TDWHD}
\widetilde{H}_{\mathrm{DW}}:=H_{\mathrm{DW}}+u^{k}\bullet \mathcal{C}_{k}^{(1)}\, ,
\eeq
where the ${u^{k}}$'s stand for Lagrange multiplier $1$-forms enforcing the primary constraint $(n-1)$-forms.  It is important to note that the total Hamiltonian is a smooth function on the polymomenta phase-space, that is, a $0$-form in $\mathcal{P}$.
 
As proposed in~\cite{IKGD}, a basic consistency requirement is the preservation of each constraint $(n-1)$-form $\mathcal{C}_{k}^{(1)}$ under the co-exterior derivative, that is, $\mathbf{d}\bullet \mathcal{C}_{k}^{(1)}\approx 0$.  Thus, 
following~(\ref{eq:evolution})  the consistency conditions are equivalent to
\beq
\label{CCD}
\{\![\widetilde{H}_{\mathrm{DW}},\mathcal{C}_{k}^{(1)}]\!\}\approx 0\, ,
\eeq
which naturally extends to the polysymplectic 
framework the analogous concept within Dirac's formalism. In a similar way, relations \eqref{CCD} can either be trivially satisfied, or impose restrictions on the Lagrange multipliers $u^{k}$'s, or they may give rise to new relations independent of the $u^{k}$'s. In the latter case, if the new relations between the polymomenta and the field variables are independent of the primary constraints \eqref{PCD}, we will term them secondary constraints, following with the terminology introduced by Dirac. If there are secondary constraints,  writing them as $(n-1)$-forms $\mathcal{C}_{k'}^{(2)}:={\mathcal{C}_{k'}}^{(2)\mu}\omega_{\mu}$, (where the subindex $k'$ runs over the appropriate set of secondary constraints), we must impose again the consistency conditions, resulting on either the fixing of the Lagrange multipliers $u^{k}$'s or obtaining new tertiary constraints. In the case we generate further constraints, we must continue so on, applying the consistency conditions  until we either fix all of the Lagrange multipliers or whenever these conditions are trivially satisfied. After the process of generating further constraints is finished, then we will have a complete set of constraint $(n-1)$-forms, that we will simply denote by $\mathcal{C}_{l}$ (the index $l$ runs over the complete set of primary, secondary, tertiary, etc., constraints), that characterizes the singular Lagrangian system. 

In particular, a Hamiltonian horizontal $p$-form $\overset{p}{F}$ is said to be first-class if its Poisson-Gerstenhaber bracket with every constraint $(n-1)$-form $\mathcal{C}_{l}$ of the system weakly vanishes, that is,
\beq
\label{FCD}
\{\![\overset{p}{F},\mathcal{C}_{l}]\!\}\approx 0\,.
\eeq
Otherwise, the Hamiltonian horizontal $p$-form $\overset{p}{F}$ is said to be second-class.
The distinction between first- and second-class 
forms allows us to separate the complete set of constraints 
$\{\mathcal{C}\}$ into subsets of first- and second-class constraint $(n-1)$-forms.  We will denote the subset of first-class constraints as $\{\mathcal{A}\}$ and the subset of second-class constraints as $\{\mathcal{B}\}$, respectively.  By considering this decomposition, we may  implement the extended De Donder-Weyl Hamiltonian, $H^{\mathrm{E}}_{\mathrm{DW}}$, 
which reads
\beq
H^{\mathrm{E}}_{\mathrm{DW}}=\widetilde{H}_{\mathrm{DW}}+{\lambda}^{i}\bullet \mathcal{A}_{i}\, ,
\eeq
where the ${\lambda}$'s  stand for horizontal $1$-forms Lagrange multipliers 
enforcing the first-class constraints, respectively, and the subindices $i$ run over the appropriate set of constraints. Within the polysymplectic framework, we will use this extended De Donder-Weyl Hamiltonian in order to obtain the correct field equations of the system, in complete analogy to the standard Dirac's formalism
for constrained theories~\cite{QSQ, Got}.

In the following section, we will implement the polysymplectic framework in order to analyze  the CMPR action described by a singular Lagrangian system.

\section{Polysymplectic formulation for the 
CMPR action}
\label{PAM}

In this section, we will analyze the CMPR model from the viewpoint of the polysymplectic formalism considering $(n-1)$-forms as our main variables. We will also introduce a  process necessary to obtain the Holst model \cite{HOLST} from the polysymplectic formulation of the CMPR action.  From now on, we will consider Lorentz signature $\sigma=-1$.

\subsection{CMPR action for gravity}\label{IM}
As mentioned before, the first BF-type formulation of General Relativity was introduced by Plebanski in~\cite{Pleb}. The main idea of this formulation was to supplement the BF theory with certain constraints, the simplicity constraints, that restrict the bivector B field as a function of the tetrad field, thus reducing Plebanski action to the self-dual formulation of General Relativity,  as discussed in~\cite{BFG, SCLQG}. Several extensions of the Plebanski action have been developed through the years. In particular, we are interested in one of such extensions named the CMPR model and introduced in~\cite{CMPR}.  The CMPR model basically consists of a formulation of General Relativity expressed as a constrained BF theory that involves the Immirzi parameter~\cite{IGRCCCG}.
In this subsection we succinctly introduce the CMPR action. We follow, as close as possible, the description of the CMPR system as formulated in \cite{MMEBFG}. To start, we will consider a $4$-dimensional background spacetime manifold $\mathcal{M}$ endowed with Minkowskian signature $\text{diag}(-1,+1,+1,+1)$ and local coordinates $\{x^{\mu}\}_{\mu=0}^{3}$. Greek letters denote spacetime coordinates while lower-case Latin letters denote internal coordinates, which can be raised and lowered with the internal Minkowski metric $\eta_{ab}=\text{diag}(-1,+1,+1,+1)$.  

The CMPR action reads
\beq
\label{CMPRA}
\hspace{-10ex}
\mathcal{S}\,[e,A,\psi,\tilde{\kappa}]:=\int_{\mathcal{M}}\left[ Q^{ab}\wedge F_{ab}-\psi_{abcd}\left(\frac{1}{2} Q^{ab}\wedge Q^{cd}+\tilde{\kappa}\left(a_{1}\eta^{acbd}-a_{2}{\epsilon}^{abcd}\right)\right) \right],
\eeq
where $F^{ab}[A]$ is the curvature 2-form 
associated to the Lorentz connection $1$-form $A$ evaluated on the 
Lie algebra $\mathfrak{so}(3,1)$, the fields $Q^{ab}[e]$ are the components of a $2$-form (the $B$-field) which depend on the tetrad $e$.  Also, $\psi$ and $\tilde{\kappa}$ stand for Lagrange multipliers enforcing certain constraints, as 
we will describe below, and the $a$'s are real parameters.  Finally, $\eta^{acbd}:=\frac{1}{2}\left(\eta^{ac}\eta^{bd}-\eta^{ad}\eta^{bc}\right)$ and ${\epsilon}^{abcd}$ are the Cartan-Killing metric and  the Levi-Civita alternating symbol, respectively.  The Lagrange multiplier 
$\psi$ fulfills the symmetry properties $\psi_{abcd}=\psi_{cdab}=\psi_{[ab][cd]}$, while the Lagrange
multiplier $\tilde{\kappa}$ is a $4$-form. 
The set of independent field variables for this action is given by the tetrad $e$, the connection $A$ and the Lagrange multipliers 
$\psi$ and $\tilde{\kappa}$.  The reason why the action~\eqref{CMPRA} describes General Relativity with Immirzi parameter and not an ordinary BF theory is 
just the fact that we are considering the
tetrad $e$ as the independent field variables, and not the 2-form $Q$, that is,  $Q^{ab}[e]$, as described for the Plebanski action in \cite{FSMLQG}. 
From our point of view, the relevance of the CMPR action is that it introduces in a natural way the Immirzi parameter into the Einstein theory.  In the context of the CMPR model, this parameter is completely determined by the configuration of the real numbers $a_{1}$ and $a_{2}$. Thus, these  real numbers play a very important role within the model as they allow us to analyze several interesting 
residual models. In particular, it has been proven that the action \eqref{CMPRA} is equivalent to several relevant models for gravity which are described in detail in \cite{MMEBFG}.

\subsection{Polysymplectic analysis}
In this section, we will apply the polysymplectic formalism to the CMPR action. 
For the reasons discussed in section~\ref{PF}, in our formulation we will consider $(n-1)$-forms
as fundamental variables. We will start by writing the Lagrangian density which defines the CMPR model as
\beq
\hspace{-12ex}
\label{CMPRL}
\mathcal{L}\left(e,A,\psi,\kappa\right)=\frac{1}{4}\epsilon^{\mu\nu\sigma\rho}F_{\mu\nu}^{ab}Q_{ab\sigma\rho}-\psi_{abcd}\left(\frac{1}{8}\epsilon^{\mu\nu\sigma\rho}Q^{ab}_{\mu\nu}Q^{cd}_{\sigma\rho}+\kappa\left(a_{1}\,\eta^{acbd}-\,a_{2}\,{\epsilon}^{abcd}\right)\right),
\eeq
where the functions $Q_{\mu\nu}^{ab}$ and $F_{\mu\nu}^{ab}$  are the coefficients associated to the differential forms $Q$ and $F$, respectively,  and $\kappa$ stands for the unique component of the volume form $\tilde{\kappa}=\kappa \omega$ that appears in the action~\eqref{CMPRA}. Following \eqref{PMD}, the polymomenta associated to the independent fields variables $e$, $A$, $\psi$ and $\kappa$ are
correspondingly  given by
\begin{subequations}\label{CMPRPM}
\begin{align}
\pi_{ab}^{\mu\nu}&:=\frac{\partial \mathcal{L}}{\partial\left(\partial_{\mu}{A_{\nu}}^{ab}\right)}=\frac{1}{2}\epsilon^{\mu\nu\sigma\rho}Q_{ab\sigma\rho} \,,\label{PM1}\\
p_{a}^{\mu\nu}&:=\frac{\partial \mathcal{L}}{\partial\left(\partial_{\mu}e_{\nu}^{a}\right)}=0\,,\label{PM2}\\
\Pi^{\mu abcd}&:=\frac{\partial \mathcal{L}}{\partial\left(\partial_{\mu}\psi_{abcd}\right)}=0\,,\label{PM3}\\ 
P^{\mu}&:=\frac{\partial \mathcal{L}}{\partial\left(\partial_{\mu}\kappa\right)}=0\,, \label{PM4}
\end{align}
\end{subequations}
thus implying that the Hessian matrix for the Lagrangian of the CMPR 
model~\eqref{CMPRL} is singular as none of the gradients of the field variables can be inverted in terms of the field variables and the polymomenta. In consequence, the definition of the polymomenta gives rise to a set of primary constraints.  We explicitly write the associated primary constraint $(n-1)$-forms as
\begin{subequations}\label{CFD}
\begin{align}
{\mathcal{C}_{\bm{A}\,}^{(1)}}_{ab}^{\nu}:&=\left(\pi_{ab}^{\mu\nu}-\frac{1}{2}\epsilon^{\mu\nu\sigma\rho}Q_{ab\sigma\rho}\right)\omega_{\mu}\approx\,0\, ,\label{CFD1}\\
{\mathcal{C}_{\bm{e}\,}^{(1)}}_{a}^{\nu}:&=p_{a}^{\mu\nu}\omega_{\mu}\approx\,0\, ,\label{CFD2}\\
{\mathcal{C}_{\bm{\psi}\,}^{(1)}}^{abcd}:&=\Pi^{\mu abcd}\omega_{\mu}\approx\,0\ ,\label{CFD3}\\
\mathcal{C}_{\bm{\kappa}\,}^{(1)}:&=P^{\mu}\omega_{\mu}\approx\,0\, .\label{CFD4}
\end{align}
\end{subequations}
where the boldface subindices for each of these constraints refer to the independent field variable from which they where obtained.
Next, by means of the covariant Legendre transformation~\eqref{DWHD}, we obtain the De Donder-Weyl Hamiltonian associated to the CMPR model, $H_{\mathrm{DW}}$, which can be written as
\beq
\hspace{-12ex}
\label{DWH}
H_{\mathrm{DW}}=-\frac{1}{2}\epsilon^{\mu\nu\sigma\rho}Q_{ab\mu\nu}{{A_{[\sigma|}}^{a}}_{c}{{A_{\rho]}}^{c\,b}}+\psi_{abcd}\left(\frac{1}{8}\epsilon^{\mu\nu\sigma\rho}Q_{\mu\nu}^{ab}Q^{cd}_{\sigma\rho}+\kappa\left(a_{1}\,\eta^{acbd}-\,a_{2}\,{\epsilon}^{abcd}\right)\right)
\eeq
with ${{A_{[\sigma|}}^{a}}_{c}{{A_{\rho]}}^{cb}}=\frac{1}{2}\left({{A_{\sigma}}^{a}}_{c}{A_{\rho}}^{c\,b}-{{A_{\rho}}^{a}}_{c}{A_{\sigma}}^{c\,b}\right)$ denoting the antisymmetrization in spacetime coordinates.
Following the polysymplectic formulation for singular Lagrangian systems of section~\ref{PF}, we may construct the total De Donder-Weyl Hamiltonian, $\widetilde{H}_{\mathrm{DW}}$, which is subject to the primary constraints \eqref{CFD},
yielding 
\beq
\label{TDWH}
\widetilde{H}_{\mathrm{DW}}=H_{\mathrm{DW}}+\lambda_{\nu}^{ab}\bullet{\mathcal{C}_{\bm{A}\,}^{(1)}}_{ab}^{\nu}+\chi_{\nu}^{a}\bullet {\mathcal{C}_{e\,}^{(1)}}_{a}^{\nu}+\xi_{abcd}\bullet{\mathcal{C}_{\bm{\psi}\,}^{(1)}}^{abcd}+\zeta\bullet{\mathcal{C}_{\bm{\kappa}\,}^{(1)}}\, ,
\eeq
where $\lambda$, $\chi$, $\xi$ and $\zeta$ are horizontal $1$-forms Lagrange multipliers enforcing the primary constraint $(n-1)$-forms. Now, we need to apply the consistency conditions on the primary constraint $(n-1)$-forms in order to obtain a complete set of constraints that characterizes the CMPR model. In particular, the consistency conditions~\eqref{CCD} applied to the constraint $(n-1)$-forms \eqref{CFD1} and \eqref{CFD2} give rise to a set of relations for the Lagrange multipliers $\chi$ and $\lambda$,
\beq
\hspace{15ex}
\epsilon^{\sigma\rho\nu\mu}Q_{c[a|\sigma\rho}{{A_{\nu}}^{c}}_{b]}-\frac{1}{2}\epsilon^{\sigma\rho\nu\mu}\frac{\partial Q_{ab\sigma\rho}}{\partial e_{\lambda}^{k}}\chi_{\nu\lambda}^{k}
&\approx & 
0   
\,,\nn\\
\left(\lambda_{[\alpha\beta]}^{a'b}+{{A_{[\alpha|}}^{a'}}_{c}{A_{\beta]}}^{cb}-\frac{1}{2}{\psi^{a'b}}_{cd}Q_{\alpha\beta}^{cd}\right)\frac{\partial\left(\epsilon^{\alpha\beta\sigma\rho}Q_{a'b\sigma\rho}\right)}{\partial e_{\mu}^{a}}
&\approx &
0
\,,
\eeq 
thus implying that these consistency conditions do not generate further constraints. However, the consistency conditions for the remaining constraint $(n-1)$-forms \eqref{CFD3} and \eqref{CFD4} generate new secondary constraints, which explicitly written as $(n-1)$-forms read
\begin{subequations}\label{CS}
\begin{align}
{\mathcal{C}_{\bm{\psi}}^{(2)}}_{\lambda}^{abcd}&:=\left(\frac{1}{8}\epsilon^{\mu\nu\sigma\rho}Q_{\mu\nu}^{ab}Q_{\sigma\rho}^{cd}+\kappa\left(a_{1}\eta^{acbd}-a_{2}\epsilon^{abcd}\right)\right)\omega_{\lambda}\, \approx\,0\, ,\label{CS1}\\
{\mathcal{C}_{\bm{\kappa}}^{(2)}}_{\lambda}&:=\psi_{abcd}\left( a_{1}\eta^{acbd}-a_{2}\epsilon^{abcd}\right)\omega_{\lambda}\approx\,0\, .\label{CS2}
\end{align}
\end{subequations}
The consistency conditions applied to these secondary constraint $(n-1)$-forms \eqref{CS} impose restrictions on the Lagrange multipliers  $\zeta$ and $\xi$, 
\beq
\frac{1}{4}\epsilon^{\mu\nu\sigma\rho}Q_{\mu\nu}^{ab}\frac{\partial Q_{\sigma\rho}^{cd}}{\partial e_{\alpha}^{k}}\chi_{\lambda \alpha}^{k}+\zeta_{\lambda}\left(a_{1}\eta ^{acbd}-a_{2}\epsilon^{abcd}\right)
& \approx &  
0
\,,\nn\\
\hspace{19ex}
\xi_{\mu abcd}\left(a_{1}\eta^{acbd}-a_{2}\epsilon^{abcd}\right)
& \approx & 
0\,,
\eeq
and hence we do not obtain further constraints. In this way we have established a complete set of constraints that characterizes the CMPR action. We are now in the position to compute the Poisson-Gerstenhaber brackets among the constraint $(n-1)$-forms of the CMPR action.
The non-vanishing brackets yield
\beq
\{\![{\mathcal{C}_{\bm{e}\,}^{(1)}}_{a}^{\mu},{\mathcal{C}_{\bm{A}\,}^{(1)}}_{bc}^{\nu} ]\!\}&=-\frac{1}{2}\frac{\partial}{\partial e_{a}^{\mu}}\left( \epsilon^{\lambda \nu \sigma \rho}Q_{cd\sigma \rho}\right)\omega_{\lambda}\, ,
\nn\\
\{\![{\mathcal{C}_{\bm{\psi}\,}^{(1)}}^{abcd},{\mathcal{C}_{\bm{\kappa}\,}^{(2)}}_\lambda ]\!\}&= \{\![ {\mathcal{C}_{\bm{\kappa}\,}^{(1)}},{\mathcal{C}_{\bm{\psi}\,}^{(2)}}_{\lambda}^{abcd}]\!\} = 
\left(a_{1}\eta^{abcd}-a_{2}\epsilon^{abcd}\right)\omega_{\lambda}\, ,  
\nn\\
\{\![ {\mathcal{C}_{\bm{e}\,}^{(1)}}_{s}^{\gamma},{\mathcal{C}_{\bm{\psi}\,}^{(2)}}_{\lambda}^{abcd} ]\!\}&=\frac{1}{4}Q_{\mu\nu}^{ab}\frac{\partial }{\partial e_{\gamma}^{s}}\left( \epsilon^{\mu\nu\sigma\rho}Q^{cd}_{\sigma\rho} \right)\omega_{\lambda}\, ,
\label{BCNV}
\eeq
while the remaining of the Poisson-Gerstenhaber brackets vanish. In consequence, we have that the full set of constraint $(n-1)$-forms that characterizes the CMPR model is
second-class,  according with the definition \eqref{FCD}, which implies that the correct field equations of the system can be obtained by means of the total De Donder-Weyl Hamiltonian \eqref{TDWH} on the constraint surface.
By simplicity, as  all the constraints 
for this model are second-class we will avoid the introduction of the notation using $\mathcal{B}$'s (as described at the end of section~\ref{PF}), keeping the 
notation with the $\mathcal{C}$'s.  
Bearing this in mind,
we will consider canonical horizontal $(n-1)$-forms as our appropriate set of field variables, such 
that their components read
\beq
{A}_{\nu\mu}^{ab}&:={A_{\nu}}^{ab}\omega_{\mu}\, , \qquad\qquad\qquad~~~ \pi^{\nu}_{ab}:=\pi^{\mu\nu}_{ab}\omega_{\mu}\, ,
\nn\\
e_{\nu\mu}^{a}&:=e_{\nu}^{a}\omega_{\mu}\, , \qquad\qquad\qquad~~~~~~~ p_{a}^{\nu}:=p_{a}^{\mu\nu}\omega_{\mu}\, ,
\nn\\
\psi_{abcd\mu}&:=\psi_{abcd}\omega_{\mu}\, , \quad\qquad\qquad~~~ \Pi^{abcd}:=\Pi^{\mu abcd}\omega_{\mu}\, ,
\nn\\
\kappa_{\mu}&:=\kappa\omega_{\mu}\, , \qquad\qquad\qquad~~~~~~~~ P:=P^{\mu}\omega_{\mu}\, .
\label{FDC}
\eeq
The De Donder-Weyl-Hamilton field equations for the $(n-1)$-forms canonical field variables associated to the CMPR model are thus given by
\beq
\mathbf{d}\bullet A_{\nu\mu}^{ab}
&=&
\{\![\widetilde{H}_{\mathrm{DW}}, A_{\nu\mu}^{ab}]\!\}=
-\lambda_{\nu\mu}^{ab}
\,,\nn\\
\mathbf{d}\bullet e_{\nu\mu}^{a}
&=& 
\{\![\widetilde{H}_{\mathrm{DW}}, e_{\nu\mu}^{a}]\!\}=
-\chi_{\nu\mu}^{a}
\,,\nn\\
\mathbf{d}\bullet \psi_{abcd\mu}
&=&
\{\![\widetilde{H}_{\mathrm{DW}},\psi_{abcd\mu}]\!\}=
\xi_{abcd\mu}
\,,\nn\\
\mathbf{d}\bullet \kappa_{\mu}
&=& 
\{\![\widetilde{H}_{\mathrm{DW}},\kappa_{\mu}]\!\}=
\zeta_{\mu}
\,,
\label{EqDWH}
\eeq
while for their corresponding polymomenta we have
\beq
\hspace{-12ex}
\mathbf{d}\bullet\pi^{\mu}_{ab}
&=& 
\{\![\widetilde{H}_{\mathrm{DW}} \,, 
\pi_{ab}^{\mu}]\!\}=
-\epsilon^{\nu\sigma\rho\mu}Q_{c[a|\nu\sigma}{{A_{\rho}}^{c}}_{b]}
\,,\nn\\
\hspace{-12ex}
\mathbf{d}\bullet p^{\mu}_{a}
&=& 
\{\![\widetilde{H}_{\mathrm{DW}}, p_{a}^{\mu}]\!\}=
\frac{1}{4}\left[2\left(\lambda_{[\alpha\beta]}^{a'b}+{{A_{[\alpha|}}^{a'}}_{c}{A_{\beta]}}^{cb}\right)-{\psi^{a'b}}_{cd}Q^{cd}_{\alpha\beta}\right]\frac{\partial}{\partial e_{\mu}^{a}}\left(\epsilon^{\alpha\beta\sigma\rho}Q_{a'b\sigma\rho}\right)
\,,\nn\\
\hspace{-12ex}
\mathbf{d}\bullet\Pi^{abcd}
&=& \{\![\widetilde{H}_{\mathrm{DW}}, \Pi^{abcd}]\!\}=
-\frac{1}{2}\left(\frac{1}{4}\epsilon^{\mu\nu\sigma\rho}Q_{\mu\nu}^{ab}Q_{\sigma\rho}^{cd}+2\kappa\left(a_{1}\eta^{acbd}-a_{2}\epsilon^{abcd}\right)\right)
\,,\nn\\
\hspace{-12ex}
\mathbf{d}\bullet P
&=& 
\{\![\widetilde{H}_{\mathrm{DW}}, P]\!\} =
-\psi_{abcd}\left(a_{1}\eta^{acbd}-a_{2}\epsilon^{abcd}\right)
\,.
\label{EqDWHpoly}
\eeq

In particular, the De Donder-Weyl-Hamilton equations~\eqref{EqDWH} fix the components of the Lagrange multipliers $1$-forms in terms of the gradients of the 
field variables as
\beq
\lambda_{\mu\nu}^{ab}
&=& 
\partial_{\mu}{A_{\nu}}^{ab}
\,,\nn\\
\chi_{\mu\nu}^{a}
&=& 
\partial_{\mu}e_{\nu}^{a}
\,,\nn\\
\xi_{\mu abcd}
&=& 
\partial_{\mu}\psi_{abcd}
\,,\nn\\
\zeta_{\mu}
&=& 
\partial_{\mu}\kappa\,.
\label{FML}
\eeq
Thus, taking into account that we have a complete set of constraints which characterizes the CMPR model, we can analyze the information which our approach gives about this system on the constraint surface, that is, on the surface where the constraints are strongly zero. Before we proceed, given the Levi-Civita and Lorentz $\mathfrak{so}(3,1)$ connections, $\Gamma^{\mu}_{\nu\lambda}$ and ${A_{\mu}}^{ab}$, respectively, we define a covariant derivative $D_{\mu}$ that fulfills the so-called tetrad 
postulate~\cite{CGR,Wald}, which simply extends the metric compatibility condition to the tetrad, and explicitly reads
\beq
\label{TP}
D_{\mu} e_{\nu}^{a}=\partial_{\mu}e_{\nu}^{a}-\Gamma^{\sigma
}_{\mu\nu}e_{\sigma}^{a}+{{A_{\mu}}^{a}}_{b}e_{\nu}^{b}=0\,.
\eeq
By considering the second-class constraints \eqref{CFD} and \eqref{CS} as strong identities we are able to manipulate the remaining De Donder-Weyl-Hamilton equations for the polymomenta~\eqref{EqDWHpoly},  obtaining the following relations 
\begin{subequations}\label{FEBF}
\begin{align}
D_{\mu}(\epsilon^{\mu\nu\sigma\rho}Q_{ab\sigma\rho})=0\label{EBF1}\, , \\
\left( F_{\mu\nu}^{ab}-{\psi^{ab}}_{cd}Q^{cd}_{\mu\nu}\right)\frac{\partial }{\partial e_{\lambda}^{k}}\left(\epsilon^{\mu\nu\sigma\rho} Q_{ab\sigma\rho}\right)=0\, , \label{EBF2}\\
\frac{1}{4}\epsilon^{\mu\nu\sigma\rho}Q_{\mu\nu}^{ab}Q_{\sigma\rho}^{cd}+2\kappa\left(a_{1}\eta^{acbd}-a_{2}\epsilon^{abcd}\right)=0\, , \label{EBF3}\\
\psi_{abcd}\left( a_{1}\eta^{acbd}-a_{2}\epsilon^{abcd}\right)=0\, . \label{EBF4}
\end{align}
\end{subequations}
These equations are exactly the field equations obtained for the CMPR model at the Lagrangian level in~\cite{CMPR} and~\cite{MMEBFG}, thus
demonstrating that the polysymplectic approach on the constraint surface is completely equivalent to the Lagrangian formulation for the model of our interest. 
Equations~\eqref{EBF1} establish that 
the quantity $\epsilon^{\mu\nu\sigma\rho}Q_{ab\sigma\rho}$ remains constant under parallel transport, that is, it is covariantly constant \cite{BFG}.  Also, from the definition of the polymomenta $\pi^{\mu\nu}_{ab}$~\eqref{PM1}, it is possible to see that the field equations~\eqref{EBF1} are analogous within our context to the Gauss law.   Equations~\eqref{EBF2}  are the most important from our point of view and shall be discussed in detail in the following paragraphs.  
Equations~\eqref{EBF3} and \eqref{EBF4} set a explicit dependence of the field $\kappa$ on 
the tetrad and a restriction on the Lagrange multiplier $\psi_{abcd}$, respectively.  

In order to analyze equations~\eqref{EBF2}, we 
will start by demonstrating the way in which the CMPR model describes gravity with an arbitrary Immirzi parameter. To this end, we start by contracting equations~\eqref{EBF3} first with the Cartan-Killing metric and then with the Levi-Civita alternating symbol, obtaining the relations
\begin{subequations}\label{EK}
\begin{align}
a_{1}\kappa\left[e\right]&=-\frac{1}{12}\left(\frac{1}{4}\epsilon^{\mu\nu\sigma\rho}Q_{\mu\nu}^{ab}Q_{ab\sigma\rho}\right)\, , \label{EK1}\\ 
a_{2}\kappa\left[e\right]&=\frac{1}{4!}\left(\frac{1}{4}\epsilon^{\mu\nu\sigma\rho}Q_{\mu\nu}^{ab}{}^{\star}Q_{ab\sigma\rho}\right)\, , \label{EK2}
\end{align}
\end{subequations}
respectively.  The ${\star}$ symbol in \eqref{EK2} denotes the Hodge dual star operator in the internal space, which should not be confused with the $*$ symbol, appearing in section \ref{PF}, that stands for the Hodge dual operator defined on the base space $\mathcal{M}$. As we are considering that, the components of the $2$-forms $Q^{ab}$ depend on the tetrad, that is, $Q^{ab}_{\mu\nu}=Q^{ab}_{\mu\nu}[e]$, equations~\eqref{EK} completely fix the field $\kappa$ in terms of the tetrad, thus implying that  $\kappa$ is not an independent dynamical variable contrary to the original
assumption, as mentioned above. Further, manipulating relations \eqref{EBF3} and \eqref{EK} give rise to the so-called simplicity constraints on $Q^{ab}_{\mu\nu}$, which explicitly read 
\begin{subequations}\label{SCBF}
\begin{align}
\epsilon^{\mu\nu\sigma\rho}Q_{\mu\nu}^{ab}Q_{\sigma\rho}^{cd}-\frac{1}{6}\eta^{acbd}\left(\epsilon^{\mu\nu\sigma\rho}Q_{\mu\nu}^{a'b'}Q_{a'b'\sigma\rho}\right)-\frac{1}{12}\epsilon^{abcd}\left(\epsilon^{\mu\nu\sigma\rho}Q_{\mu\nu}^{a'b'}{}^{*}Q_{a'b'\sigma\rho}\right)&=0\, ,\\
2a_{2}\epsilon^{\mu\nu\sigma\rho}Q_{\mu\nu}^{ab}Q_{ab\sigma\rho}+a_{1}\epsilon^{\mu\nu\sigma\rho}Q_{\mu\nu}^{ab}{}^{*}Q_{ab\sigma\rho}&=0\, .
\end{align}
\end{subequations}
As discussed in \cite{CMPR}, these simplicity constraints impose restrictions on the functions $Q^{ab}_{\mu\nu}$ and, for the non-degenerate case ($Q^{ab}_{\mu\nu}\neq 0$), have a unique solution which may be written in terms of the tetrad as
\beq
\label{SQ}
Q_{\mu\nu}^{ab}=\left( \alpha {\epsilon^{ab}}_{cd}+\beta \delta_{cd}^{ab}\right)e_{[\mu}^{c}e_{\nu]}^{d}\, ,
\eeq
where $\delta_{cd}^{ab}:=(1/2)(\delta^a_c \delta^b_d-\delta^a_d \delta^b_c)$, while $\alpha$ and $\beta$ are real numbers subject to the condition
\beq
\label{RCS}
\frac{a_{2}}{a_{1}}=\frac{\alpha^{2}-\beta^{2}}{4\alpha\beta}\, .
\eeq
The solution \eqref{SQ} for the simplicity constraints \eqref{SCBF} is discussed in detail in references~\cite{CMPR} and~\cite{MMEBFG} where, in particular, it is shown that for $\alpha,\ \beta\neq 0$, the parameter $\gamma:=\beta/\alpha$ is related to the inverse of the Immirzi parameter. 
At this point it is important to mention that the solution \eqref{SQ} of the simplicity constraints \eqref{SCBF} is 
a crucial step in order to obtain General Relativity from the BF formalism of our interest.  Indeed,  substituting~\eqref{SQ}  into the original action \eqref{CMPRA} we obtain the action proposed by Holst~\cite{HOLST}.  This was the 
original idea behind the Plebanski formulation of General Relativity \cite{Pleb}: the characteristic topological behavior of a BF theory is broken by introducing the simplicity constraints, and thus we recover a well-defined gravity model~\cite{BFG, CMPR, MMEBFG, SLNP}. In that respect, we note that substituting solution \eqref{SQ} into the field equation \eqref{EBF1}
together with the identity
\beq
\label{F1}
ee_{a}^{[\mu}e_{b}^{\nu]}=\frac{1}{4}\epsilon_{abcd}\epsilon^{\mu\nu\sigma\rho}e_{[\sigma}^{c}e_{\rho]}^{d}\, ,
\eeq
where $e$ is the determinant of the tetrad, it is possible to write the field equations~\eqref{EBF1} in a more familiar way, namely
\beq
\label{CM}
D_{\mu}\left( e\,e^{[\mu}_{a}e^{\nu]}_{b}\right)&=0\, ,
\eeq
which in the context of differential geometry stands for the well known metric compatibility condition. This condition emerges naturally either within the spin connection formulation of the Holst actions \cite{HOLST} or with the De Donder-Weyl formulation for vielbein gravity \cite{IKHFVG, DVVG}.

Now we are in the position to discuss the field equations \eqref{EBF2}. We initiate by noting that the field equations obtained by varying the CMPR action with respect to the 2-forms $Q^{ab}$ at the Lagrangian level in \cite{MMEBFG} are very similar to the relations \eqref{EBF2}, the difference being
that, within our approach, these field equations naturally include the projector term $\partial \left(\epsilon^{\mu\nu\sigma\rho} Q_{ab\sigma\rho}\right)/ \partial e_{\lambda}^{k}$.  
This term emerged as we are considering  from the beginning variations of the action \eqref{CMPRA} with respect to the tetrad, together with the explicit dependence 
$Q^{ab}[e]$.  This permitted us to obtain Einstein equations in a natural way, after solving the simplicity constraints of the theory, as suggested in \cite{FSMLQG}. Indeed, by substituting the field equations \eqref{EBF3} and~\eqref{EBF4}, subject to relations \eqref{EK} and~\eqref{SQ}, into the field equations \eqref{EBF2}, it is possible to directly rewrite them as
\beq
\label{EBF2R}
F_{\mu\nu}^{ab}\left(\delta^{cd}_{ab}-\gamma\,{\epsilon^{cd}}_{ab}\right)\frac{\partial}{\partial e_{\lambda}^{k}}\left( e e^{[\mu}_{c}e^{\nu]}_{d}\right)=0\, .
\eeq
Further, by means of identity \eqref{F1}, we may identify  the derivative term in \eqref{EBF2R} as
\beq
\label{PS}
\frac{\partial}{\partial e_{\lambda}^{k}}\left(e\,e^{[\mu}_{c}e^{\nu]}_{d}\right)=\frac{3!e}{2} e_{c}^{[\mu}e_{d}^{\nu}e_{k}^{\lambda]}\, ,
\eeq
which is nothing but the projector implemented in \cite{SLNP} in order to write explicitly the Einstein equations from the BF field equations associated to the CMPR action obtained at the Lagrangian level in \cite{CMPR, MMEBFG}. We would like to emphasize that, within our approach, this projector appears in
a completely natural manner. Finally, in light of the first Bianchi identity, $R_{\mu[\nu\sigma\rho]}=0$, where $R_{\mu\nu\sigma\rho}=e_{\mu}^{a}e_{\nu}^{b}F_{\sigma\rho ab}$ stands for the Riemann curvature tensor defined over the base space manifold $\mathcal{M}$, jointly with relation \eqref{PS}, it is possible to write the field equations \eqref{EBF2R} in a more familiar way, namely, 
\beq
\label{EE}
F_{\mu\nu}^{ab}e_{a}^{\lambda}e_{b}^{\nu}e_{k}^{\mu}-\frac{1}{2}R\,e_{k}^{\lambda}=0\, ,
\eeq
where $R$ stands for the scalar curvature associated to base space manifold $\mathcal{M}$. These
equations are simply the well known Einstein equations~\cite{SLNP}.

In summary,  while in the standard Lagrangian formulation of the CMPR model the metric compatibility condition \eqref{CM} and the Einstein equations \eqref{EE} are obtained either 
recursively through the Holst action or by introducing an 
appropriate projector as discussed in~\cite{SLNP}, 
within the polysymplectic framework both~\eqref{CM} and~\eqref{EE}
are directly realized as a consequence of the 
field equations without need to invoke the Holst 
action at all.  Further, the polysymplectic framework also incorporates the projector~\eqref{PS} from first principles, thus 
recovering Einstein equations in a succinct manner.

\subsection{From the CMPR action to the Holst action}
In this section, we will describe the way in which the polysymplectic formulation of the CMPR action reduces into the Holst action.  
We will closely follow the Lagrangian analysis of the CMPR action as presented in \cite{MMEBFG}.

To start, we briefly review the situation at the Lagrangian level.  As discussed in \cite{BFG}, the introduction of the 
solution~\eqref{SQ} to the simplicity constraints \eqref{SCBF} into the CMPR action collapses it into the Holst action. In particular, it is possible to see that, after solving the simplicity constraints \eqref{SCBF} on $Q^{\mu\nu}_{ab}$ and by replacing the solution \eqref{SQ} into the original action \eqref{CMPRA}, the remaining action only depends on the tetrad and the spin connection as independent variables.  Hence, for 
$\alpha\neq 0$, this action reduces to 
\beq
\label{HAR}
\mathcal{S}\,[e,A]=\frac{\alpha}{2}\int_{\mathcal{M}}d^{4}x\,e\,e^{\mu}_{a}e^{\nu}_{b}\left(\delta^{ab}_{cd}-\gamma\,{\epsilon^{ab}}_{cd}\right)F^{cd}_{\mu\nu}\, ,
\eeq
which introduces the parameter $\gamma=\beta/\alpha$, as described above.  Thus, the Lagrangian 
formulations starts by considering the standard variational procedure for the Holst action~\eqref{HAR}.

Nevertheless, within the polysymplectic approach, the variational procedure at the Lagrangian level may be 
seen equivalently by replacing the solution
to the simplicity constraints~\eqref{SQ} at every stage within the polysymplectic formulation of the CMPR action as far as we restrict each step to the surface where the field variables $\psi_{abcd}$ and $\kappa$, together with their associated polymomenta,  are
held as strongly zero.  This is a consequence 
that the physical information related with these polymomenta phase-space variables is already encoded in the solution~\eqref{SQ}. To see this, we will explicitly obtain Holst action from the 
CMPR action at the polysymplectic level.
From now on, we will use a supraindex $\holst$ to denote this process within the polysymplectic analysis.   Hence, we see that the set of polymomenta \eqref{CMPRPM} reduces to
\beq
{\pi^{\holst}}_{ab}^{\mu\nu}
&=& 
\left( \alpha \delta_{ab}^{cd}-\beta {\epsilon^{cd}}_{ab}\right)ee^{[\mu}_{c}e^{\nu]}_{d}
\,,\nn\\
{p^{\holst}}_{a}^{\mu\nu}
&=& 0 \,,
\label{PM1R}
\eeq
where these polymomenta are exactly the ones associated to the spin connection, ${A_{\mu}}^{ab}$, and the tetrad, $e_{\mu}^{a}$, which may be obtained 
from~\eqref{HAR}, respectively. These polymomenta also imply that, the set of primary constraint $(n-1)$-forms \eqref{CFD} becomes
\beq
{\mathcal{C}^{(1)\holst}_{\bm{A}\,}}_{ ab}^{\nu}
&=& 
\left({\pi^{\holst}}_{ab}^{\mu\nu}-\left( \alpha \delta_{ab}^{cd}-\beta {\epsilon^{cd}}_{ab}\right)e e^{[\mu}_{c}e^{\nu]}_{d}\right)\omega_{\mu}\approx\,0
\,, \nn \\
{\mathcal{C}^{(1)\holst}_{\bm{e}\,}}_{a}^{\nu}
&=&
{p^{\holst}}_{a}^{\mu\nu}\omega_{\mu}\approx\,0\, .
\label{CSE1R}
\eeq

Besides, by means of~\eqref{SQ}, the De Donder-Weyl Hamiltonian \eqref{DWH} for the Holst action reads
\beq
\label{DWHR}
{H^{\holst}_{\mathrm{DW}}}(e, A, p, \pi)=-\left( \alpha \delta_{ab}^{cd}-\beta {\epsilon^{cd}}_{ab}\right)ee^{[\mu}_{c}e^{\nu]}_{d}{{A_{[\mu|}}^{a}}_{c'}{A_{\nu]}}^{c'b}\, .
\eeq
In consequence, the total De Donder-Weyl Hamiltonian for the Holst action is given by
\beq
\label{TotalDWHR}
{\widetilde{H}^{\holst}_{\mathrm{DW}}}(e, A, p, \pi)={H^{\holst}_{\mathrm{DW}}}+{\lambda^{\holst}}_{\nu}^{ab}\bullet{\mathcal{C}^{(1)\holst}_{\bm{A}\,}}_{ ab}^{\nu}+{\chi^{\holst}}_{\nu}^{a}\bullet{\mathcal{C}^{(1)\holst}_{e\,}}_{a}^{\nu}\, ,
\eeq
where ${\lambda^{\holst}}_{\nu}^{ab}$ and ${\chi^{\holst}}_{\nu}^{a}$ stand for Lagrange multiplier $1$-forms enforcing the primary constraint $(n-1)$-forms~\eqref{CSE1R}. 
One may directly show that the consistency conditions \eqref{CCD} applied to the primary constraint $(n-1)$-forms \eqref{CSE1R} do not generate further constraints and, thus, the corresponding De Donder-Weyl-Hamilton equations for the Holst action are given by 
\beq
\hspace{-13ex}
\mathbf{d}\bullet {A_{\nu\mu}}^{ab}
&=&
\{\![{\widetilde{H}^{\holst}_{\mathrm{DW}}}, A_{\nu\mu}^{ab}]\!\}=
-{\lambda^{\mathrm{\mathbf{H}}}}_{\nu\mu}^{ab}
\,,\nn\\
\hspace{-13ex}
\mathbf{d}\bullet e_{\nu\mu}^{a}
&=&
\{\![{\widetilde{H}^{\holst}_{\mathrm{DW}}}, e_{\nu\mu}^{a}]\!\}=
-{\chi^{\mathrm{\mathbf{H}}}}_{\nu\mu}^{a}
\,,\nn\\
\hspace{-13ex}
\mathbf{d}\bullet{\pi^{\holst}}_{ab}^{\mu}
&=& 
\{\![{\widetilde{H}^{\holst}_{\mathrm{DW}}}, {\pi^{\holst}}_{ab}^{\mu}]\!\}=
-2\alpha\left(\delta_{c[a|}^{c'd'}-\gamma{\epsilon^{c'd'}}_{c[a|}\right)e\, e_{c'}^{[\rho}e_{d'}^{\mu]}{{A_{\rho}}^{c}}_{|b]}
\,,\nn\\
\hspace{-13ex}
\mathbf{d}\bullet {p^{\holst}}_{a}^{\mu}
&=&
\{\![{\widetilde{H}^{\holst}_{\mathrm{DW}}}, {p^{\holst}}_{a}^{\mu}]\!\}=
\alpha\left({\lambda^{\mathrm{\mathbf{H}}}}_{[\rho\sigma]}^{a'b}+ {{A_{[\rho|}}^{a'}}_{c'}{A_{\sigma]}}^{c'b}\right)\left(\delta_{a'b}^{cd}-\gamma{\epsilon^{cd}}_{a'b}\right)\frac{\partial}{\partial e_{\mu}^{a}}\left(e e_{c}^{[\rho}e_{d}^{\sigma]}\right)  .
\label{DWHER}
\eeq
As before, the first two equations only fix the
Lagrange multipliers in a totally equivalent way as in the first two formulae in~\eqref{FML}, while the other two equations give rise to the metric compatibility condition \eqref{CM} and the Einstein equations \eqref{EBF2R}, respectively, on the surface where the constraints \eqref{CSE1R} are strongly zero.  
These results are completely consistent with the Lagrangian analysis of the Holst action as performed, for example, in~\cite{HOLST}. Finally, it is possible to realize that, the polymomenta \eqref{PM1R} and the corresponding total De Donder-Weyl Hamiltonian \eqref{TotalDWHR} are
entirely identical  to those obtained in \cite{IKHFVG} and \cite{DVVG}. 
This demonstrates that, within the polysymplectic framework, it is possible to bring the CMPR action into the Holst action in a  consistent manner.

\section{Conclusions}\label{SC}

Within the polysymplectic framework for classical field theories we analyzed the CMPR action, which stands for a BF-type formulation for General Relativity involving an
arbitrary Immirzi parameter. The CMPR model is described by a singular Lagrangian system, and thus we implemented the scarcely exploited polysymplectic extension of Dirac’s formalism for constrained singular Lagrangian systems as introduced in \cite{IKGD,IKHFVG}. For the case of our interest, implementation of this polysymplectic extension resulted completely successful as it allowed us a straight interpretation of the constraint content of the model and lead us to the precise field equations. Particularly, we found that the CMPR model is entirely characterized by a set of second-class constraints.
After solving the simplicity constraints of the theory, the Einstein equations emerged in a 
natural manner as within the  polysymplectic formalism the De Donder-Weyl Hamilton field equations already included 
a suitable projector term that is required in order to straightforwardly extract 
Einstein equations from the CMPR action.   Additionally, as we exposed in the previous section, by considering the 
restriction of polymomenta phase space to the constraint surface 
defined by the second-class constraints, and by replacing the solution to the simplicity constraints into the polysymplectic analysis of the CMPR action, we were able to recover the polysymplectic formulation of the Holst action. 
We must emphasize that this reduction procedure was implemented directly at every structure of the polysymplectic formalism for the CMPR model, and 
not by reducing the CMPR action to the Holst 
action from the beginning, as in the Lagrangian 
formulation.
All these results effectively demonstrated that the polysymplectic extension of Dirac's algorithm for constrained systems is an adequate candidate for the analysis of physically motivated singular Lagrangians.

Further, as within the case of  standard Dirac analysis for singular systems, one may wonder about the connection between constraints and gauge symmetries within the polysymplectic approach developed here.
Indeed, for the physically motivated models analyzed within the polysymplectic formalism in the literature, all the Lagrange multipliers are fixed, thus avoiding the presence of first-class constraints. In consequence, 
gauge transformations may be analyzed carefully through the so-called covariant momentum map. 
This  map corresponds to an extension and refinement of  gauge invariance  on the corresponding polymomenta phase space~\cite{Gotay} and satisfies Noether's theorem by including all the gauge information of a classical field theory, enabling us to define the notion of first class constraints as generators of gauge symmetries on the instantaneous phase space, that is, once an initial Cauchy surface has been singled out. However, potential relations between the covariant momentum map with the polysymplectic formalism are poorly explored. In particular, relations among the constraints at the polysymplectic level and the generators of gauge symmetries are not 
completely understood.   
Therefore, it is mandatory to get a better understanding of both  structures in order to develop a complete covariant counterpart to the usual Hamiltonian formalism. This will be done elsewhere.


\section*{Acknowledgments}
The authors would like to acknowledge financial support from CONACYT-Mexico
under projects CB-2014-243433 and CB-2017-283838.

\end{document}